\documentclass[a4paper,12pt]{article}
\usepackage{amssymb,amsmath,amsfonts}
\usepackage{fullpage}
\usepackage[latin1]{inputenc}
\newcommand{\diag}{ {\rm diag}}
\begin{document}
\title{Improved estimators for dispersion models with dispersion covariates}
\date{}
\author{Alexandre B. Simas$^{a,}$\footnote{Corresponding author. E-mail: alesimas@impa.br}, Andréa V. Rocha$^{b,}$\footnote{E-mail: andrea@de.ufpb.br} ~ and Wagner Barreto-Souza$^{c,}$\footnote{E-mail: wagnerbs85@hotmail.com}\\\\
\centerline{\small{
$^a$Associa\c{c}\~ao Instituto Nacional de Matem\'atica Pura e Aplicada, IMPA,}}\\
\centerline{\small{
Estrada D. Castorina, 110, Jd. Bot\^anico, 22460-320, Rio de Janeiro-RJ, Brasil}}\\
\centerline{\small{
$^b$Departamento de Estat\' \i stica, 
Universidade Federal da Para\' iba,
}}\\
\centerline{\small{
Cidade Universitária - Campus I, 58051-970, Jo\~ ao Pessoa-PB, Brasil}}\\
\centerline{\small{
$^c$Departamento de Estat\' \i stica, 
Universidade de S\~ao Paulo,
}}\\
\centerline{\small{
Rua do Mat\~ao, 1010, 05508-090, S\~ao Paulo-SP, Brasil}}}
\sloppy
\maketitle
\begin{abstract}
In this paper we discuss improved estimators for the regression and the dispersion parameters in
an extended class of dispersion models (J\o rgensen, 1996). This class extends the regular dispersion
models by letting the dispersion parameter vary throughout the observations, and contains the dispersion
models as particular case. General formulae for the second-order bias are obtained explicitly in dispersion
models with dispersion covariates, which generalize previous results by Botter and Cordeiro (1998), Cordeiro and McCullagh (1991), Cordeiro and Vasconcellos (1999), and Paula (1992). The practical use of the formulae is that we can derive closed-form
expressions for the second-order biases of the maximum likelihood estimators of the regression and dispersion
parameters when the information matrix has a closed-form. Various expressions for the second-order biases
are given for special models. The formulae have advantages for numerical purposes because they require only
a supplementary weighted linear regression. We also compare these bias-corrected estimators
with two different estimators which are also bias-free to the second-order that are based on bootstrap methods.
These estimators are compared by simulation.
\end{abstract}

\section{Introduction}

The class of dispersion models was introduced by J\o rgensen (1997a) and represents a collection of probability
density functions which has as particular cases, the proper dispersion models also introduced by J\o rgensen (1997b), 
and the well-known one parameter exponential family. It is possible to introduce a regression structure, and that is what will be done in this work. We
also allow a regression structure on the dispersion parameter. Thus, this regression structure generalizes the exponential
family nonlinear models (Cordeiro and Paula, 1989), the generalized linear models with dispersion covariates
(see, for instance, Botter and Cordeiro, 1998), and the generalized linear models (McCullagh and Nelder, 1989).
We will call from now on, the dispersion model together with its regression structure simply by dispersion model. Recently, Simas et al. (2009b) studied asymptotic tail properties of distributions in the class of dispersion models.

Few attempts have been made to develop second-order asymptotic theory for dispersion models in order to have better likelihood inference procedures. An asymptotic formula of order $n^{-1/2}$, where $n$ is the sample size, for the skewness of the distribution of $\hat{\beta}$ in  dispersion models was obtained by Simas et al. (2009c). Moreover, Rocha et al. (2009) obtained a matrix expression for the covariance matrix up to the second-order for dispersion models with this regression structure.

The problem of modeling variances has been largely discussed in the statistical literature particularly in the econometric area (see, for instance, Harvey, 1976).
Under normal errors, Atkinson (1985) present some graphical methods to detect heteroscedasticity. Moving
away from normal errors, Smyth (1989) describes a method which allows modeling the dispersion parameter in some generalized linear models.

It is known that maximum likelihood estimators (MLEs) in nonlinear regression models are generally biased. These bias become
problematic when the study is being done in small samples. Bias does not pose a serious problem when the sample size $n$
is large, since its order is typically ${\cal O}(n^{-1})$, whereas the asymptotic standard error has order ${\cal O}(n^{-1/2})$.
Several authors have explored bias in regression models. Pike et al. (1979) investigated the magnitude of the bias in
unconditional estimates from logistic linear models, when the number of strata is large. Cordeiro
and McCullagh (1991) gave a general bias formulae in matrix notation for generalized linear models. Furthermore, Simas et al. (2009a) obtained matrix expressions for the second-order bias of the MLEs in a general beta regression model.

The method used to obtain expressions of the ${\cal O}(n^{-1})$ bias of the parameters of this class of dispersion 
models is the one given by Cox and Snell (1968). It is also possible to perform bias adjustment using the estimated bias from a bootstrap resampling scheme, which requires no explicit derivation of the bias function.

The chief goal of this paper is to obtain closed-form expressions for the second order biases of the MLEs
of the parameters, of the means of the responses, and of the precision parameters of the model.
The results are used to define bias corrected estimators to order ${\cal O}(n^{-1})$. We also consider bootstrap bias
adjustment.

The rest of this paper unfolds as follows. In Section 2, we introduce the class of dispersion models with
dispersion covariates along with the score function and Fisher's information matrix. In Section 3, we derive a matrix expression for the second order biases of the MLEs of the parameters, and
consider analytical and bootstrap bias correction schemes. We also show how the
biases of the MLEs of the parameters can be easily computed by means of auxiliary weighted linear regressions.
In Section 4, we obtain the second order biases of the MLEs of the means of the responses and precision parameters
of the model.  In Section 5, we consider some special cases in detail. In Section 6, we present simulation results that show that the proposed estimators have better performance
in small samples, in terms of bias, than the original MLEs. 
Finally, the paper is concluded in Section 7 with some final remarks. In the Appendix we give explicit expressions for the quantities needed
to calculate the ${\cal O}(n^{-1})$ bias of the MLEs of the parameters.
\section{Dispersion models with dispersion covariates}

Let the random variables $Y_1,\ldots,Y_n$ be independent with each $Y_i$ having a probability density
function of the form
\begin{equation}\label{dens}
\pi(y;\mu_i,\phi_i) = {\rm exp}\{\phi t(y,\mu_i) + a(\phi_i,y)\},\quad y\in\mathbb{R},
\end{equation}
where $a(\cdot,\cdot)$ and $t(\cdot,\cdot)$ are given functions, $\phi>0$ and $\mu$ varies in an interval
of the line. Exponential dispersion models are a special case of (\ref{dens}), obtained by taking
$t(y,\mu) = \theta y-b(\theta)$, where $\mu = b'(\theta)$. Proper dispersion models are also a special
case of (\ref{dens}), obtained by taking $a(\phi,y) = d_1(\phi) + d_2 (y)$, where $d_1(\cdot)$ and
$d_2(\cdot)$ are known functions. If $Y$ is continuous, $\pi(\cdot)$ is assumed to be a density with
respect to Lebesgue measure, while if $Y$ is discrete $\pi(\cdot)$ is assumed to be a density with respect
to counting measure. We call $\phi$ the precision parameter and $\sigma^2= \phi^{-1}$ the dispersion parameter.
Similarly, the parameter $\mu$ may generally be interpreted as a kind of location parameter, but $\mu$
is not generally the expectation of the distribution.

In order to introduce a regression structure in the class of models (\ref{dens}), we assume that
\begin{equation}\label{regr}
g_1(\mu_i) = \eta_{1i} = f_1(x_i^T;\beta)\hbox{~~and~~} g_2(\phi_i) = \eta_{2i} = f_2(z_i^T;\theta),\quad i=1,\ldots,n,
\end{equation}
where $x_i = (x_{i1},\ldots,x_{im_1})^T$ and $z_i = (z_{1i},\ldots,z_{im_2})$ are $m_1$ and $m_2$-vectors of
nonstochastic independent variables associated with the $i$th response which need not to be exclusive, $\beta=(\beta_1,\ldots,\beta_p)^T$ is a
$p$-vector of unknown parameters, $\theta = (\theta_1,\ldots,\theta_q)^T$ is a $q$-vector of unknown parameters,
$g_1(\cdot)$ and $g_2(\cdot)$ are strictly monotonic and twice continuously differentiable and are usually referred
to as link functions, $f_1(\cdot;\cdot)$ and $f_2(\cdot;\cdot)$ are, possibly nonlinear, twice continuously
differentiable functions with respect to $\beta$ and $\theta$, respectively. The regression parameters $\beta$ and $\theta$
are assumed to be functionally independent. The regression structures link
the covariates $x_i$ and $z_i$ to the parameters of interest $\mu_i$ and $\phi_i$, respectively, where $\mu_i$,
as described above, is not necessarily the mean of $Y_i$. The $n\times p$ matrix of derivatives of $\eta_1$
with respect to $\beta$ is denoted by $\tilde{X} = \tilde{X}(\beta) = \partial\eta_1/\partial\beta$, and
the $n\times q$ matrix of derivatives of $\eta_2$ with respect to $\theta$ is denoted by $\tilde{Z} = \tilde{Z}(\theta) = \partial\eta_2/\partial\theta$,
and these matrices are assumed to have ranks $p$ and $q$ for all $\beta$ and all $\theta$, respectively. It
is also assumed that the usual regularity conditions for maximum likelihood estimation and large sample
inference hold; see Cox and Hinkley (1974, Chapter 9).

Consider a random sample $y_1,\ldots,y_n$ from (\ref{dens}). The log-likelihood function for this class of dispersion models with dispersion covariates has the form
\begin{equation}\label{loglik}
\ell(\beta,\theta) = \sum_{i=1}^n \{ \phi_i t(y_i,\mu_i) + a(\phi_i,y_i)\},
\end{equation}
$\mu_i = g_1^{-1}(\eta_{1i})$, $\phi_i = g_2^{-1}(\eta_{2i})$, as defined in (\ref{regr}), are functions of $\beta$ and $\theta$, respectively.

The components of the score vector, obtained by differentiation of the log-likelihood function with respect to the parameters, are given, for
$r=1,\ldots,p$, as
$$U_r(\beta,\theta) = \frac{\partial\ell(\beta,\theta)}{\partial\beta_r} = \sum_{i=1}^n \phi_i t'(y_i,\mu_i) \frac{d\mu_i}{d\eta_{1i}}\frac{\partial\eta_{1i}}{\partial\beta_r},\quad r=1,\ldots,p,$$
where $t'(y_i,\mu_i) = \partial t(y_i,\mu_i)/\partial\mu_i$, and for $R = 1,\ldots,q$
$$U_R(\beta,\theta) = \frac{\ell(\beta,\theta)}{\partial\theta_R} = \sum_{i=1}^n \{t(y_i,\mu_i) + a'(\phi_i,y_i)\}\frac{d\phi_i}{d\eta_{2i}}\frac{\partial\eta_{2i}}{\partial\theta_R},\quad R=1,\ldots,q,$$
where $a'(\phi_i,y_i) = \partial a(\phi_i,y_i)/\partial \phi_i$. Further, the regularity conditions implies that
$$E\left( t'(y_i,\mu_i)\right) = 0\hbox{~~and~~} E\left(t(y_i,\mu_i)\right) = - E\left(a'(\phi_i,y_i)\right).$$

Let $d_{ri} = E\left( \partial^r t(y_i,\mu_i)/\partial\mu_i^r\right)$ and $\alpha_{ri} = E\left( \partial^r a(\phi_i,y_i)/\partial \phi_i^r\right)$, note
that $d_1 = 0$, $d_0 = - \alpha_1$, further,
let $t^\ast = (t'(y_1,\mu_1),\ldots,t'(y_n,\mu_n))^T$, $v = t(y_i,\mu_i) + a'(\phi_i,y_i)$, also, define the matrix \label{formulat1t2} $T_1 = {\rm diag} (d\mu_i/d\eta_{1i})$,
$T_2 = {\rm diag} (d\phi_i/d\eta_{2i})$, $\Phi = {\rm diag}(\phi_i)$, with ${\rm diag}(\mu_i)$ denoting the $n\times n$ diagonal matrix with typical
element $\mu_i$, $i=1,\ldots,n$. Therefore, we can write the $(p+q)\times 1$ dimensional score vector $U(\zeta)$ in the form
$(U_\beta(\beta,\theta)^T,U_\theta(\beta,\theta)^T)^T$, with
\begin{equation}\label{scorevec}
\begin{array}{ccc}
U_\beta(\beta,\theta) & = &\tilde{X}^T \Phi T_1 t^\ast,\\
U_\theta(\beta,\theta) & =&\tilde{Z}^T T_2 v.
\end{array}
\end{equation}
The MLEs of $\beta$ and $\theta$ are obtained as the solution of the nonlinear system $U(\zeta) = 0$. In practice, the MLEs can be obtained through a numerical
maximization of the log-likelihood function using a nonlinear optimization algorithm, e.g., BFGS. For details, see Press et al. (1992).

It is possible to obtain Fisher's information matrix for the parameter vector $\zeta = (\beta^T,\theta^T)^T$ as
$$K(\zeta) = \left(\begin{array}{cc}
K_{\beta}(\zeta) & 0 \\
0 & K_{\theta}(\zeta)
\end{array}\right),
$$
where, $K_{\beta}(\zeta) = \tilde{X}^T \Phi W_\beta \tilde{X}$, $K_{\theta}(\zeta) = \tilde{Z}^T W_\theta \tilde{Z}$,\label{matrizes} $W_\beta = {\rm diag} \left(-d_{2i} (d\mu_i/d\eta_{1i})^2\right)$
and $W_\theta = {\rm diag} \left(-\alpha_{2i} (d\phi_i/d\eta_{2i})^2\right)$. Further, note that the parameters $\beta$ and $\theta$ are globally
orthogonal (Cox and Reid, 1987). Furthermore, the MLEs $\hat{\zeta}$ and $K(\zeta)$ are consistent estimators of $\zeta$ and $K(\zeta)$, respectively,
where $K(\hat{\zeta})$ is the Fisher's information matrix evaluated at $\hat{\zeta}$. Assuming that $J(\zeta) = \lim_{n\to\infty} K(\zeta)/n$ exists
and is nonsingular, we have that $\sqrt{n}\left(\hat{\zeta}-\zeta\right) \stackrel{d}{\to} N_{p+q}(0, J(\zeta)^{-1})$, where, $\stackrel{d}{\to}$
denotes convergence in distribution. Hence, if $\zeta_r$ denotes the $r$th component of $\zeta$, it follows that
$$\left(\hat{\zeta}-\zeta \right) \{K(\hat{\zeta})^{rr}\}^{-1/2} \stackrel{d}{\to} N(0,1),$$
where $K(\hat{\zeta})$ is the $r$th diagonal element of $K(\hat{\zeta})^{-1}$. Then, if $0<\alpha<1/2$, and $q_\gamma$ represents the $\gamma$ quantile
of the $N(0,1)$ distribution, we have, for $r = 1,\ldots, p$, $\hat{\beta}_r \pm q_{1-\alpha/2}\left( K_\beta(\hat{\zeta})^{rr} \right)^{1/2}$ and
$\hat{\theta}_R \pm q_{1-\alpha/2} \left(K_\theta(\hat{\zeta})^{RR}\right)^{1/2}$ as the limits of asymptotic confidence intervals for $\beta_r$ and
$\theta_R$, respectively, both with asymptotic coverage of $100(1-\alpha)\%$, where $K_\beta(\hat{\zeta})^{rr}$ is the $r$th diagonal element of
$K_\beta(\hat{\zeta})^{-1}$ and $K_\theta(\hat{\zeta})^{RR}$ is the $R$th diagonal element of $K_\theta(\hat{\zeta})^{-1}$. The asymptotic variances
of $\hat{\beta}_r$ and $\hat{\theta}_R$ are estimated by $K_\beta(\hat{\zeta})^{rr}$ and $K_\theta(\hat{\zeta})^{RR}$, respectively.

\section{Bias correction of the MLEs of $\beta$ and $\theta$}

We begin by obtaining an expression for the second order biases of the MLEs of $\beta$ and $\theta$ in this class of dispersion models with
dispersion covariates using Cox and Snell's (1968) general formula. With this expression we will be able to obtain bias corrected estimates of the
unknown parameters.

We now introduce the following total \label{cumulants} log-likelihood derivatives in which we reserve lower-case subscripts $r,s,t,u,\ldots$ to denote components
of the $\beta$ vector and upper-case subscripts $R,S,T,U,\ldots$ for components of the $\theta$ vector: $U_r = \partial \ell/\partial \beta_r$,
$U_{rS} = \partial^2\ell/\partial\beta_r\theta_S$, $U_{rsT} = \partial^3\ell/\partial\beta_r\partial\beta_s\partial\theta_T$, and so on. The standard
notation will be adopted for the moments of the log-likelihood derivatives: $\kappa_{rs} = E(U_{rs})$, $\kappa_{r,s} = E(U_rU_s)$,
$\kappa_{rs,T} = E(U_{rs}U_T)$, etc., where all $\kappa$'s to a total over sample and are, in general, of order ${\cal O}(n)$. We define the derivatives
of the moments by $\kappa_{rs}^{(t)} = \partial \kappa_{rs}/\partial\beta_t$, $\kappa_{rs}^{(T)} = \partial\kappa_{rs}/\partial\theta_T$, etc. Not
all the $\kappa$'s are functionally independent. For example, $\kappa_{rs,t} = \kappa_{rst} - \kappa_{rs}^{(t)}$ gives the covariance between the first
derivative of $\ell(\beta,\theta)$ with respect to $\beta_t$ and the mixed second derivative with respect to $\beta_r,\beta_s$. Further,
let $\kappa^{r,s} = - \kappa^{rs}$, $\kappa^{R,s} = -\kappa^{Rs}$, $\kappa^{r,S} = -\kappa^{rS}$ and $\kappa^{R,S} = -\kappa^{RS}$ be typical elements
of $K(\zeta)^{-1}$, the inverse of the Fisher's information matrix, which are ${\cal O}(n^{-1})$.

Let $B(\hat{\beta}_a)$ and $B(\hat{\theta}_A)$ be the ${\cal O}(n^{-1})$ bias of the MLEs for the $a$th component of the
parameter vector $\beta$ and the $A$th component of the parameter vector $\theta$, respectively. From the general
expression for the multiparameter ${\cal O}(n^{-1})$ biases of the MLEs given by Cox and Snell (1968), and from
the global orthogonality of the parameter (see details in the Appendix), we can write
\begin{equation}\label{biasbeta}
B(\hat{\beta}_a) = \sum_{r,s,u} \kappa^{ar} \kappa^{su} \left\{\kappa_{rs}^{(u)} -\frac{1}{2} \kappa_{rsu}\right\},
\end{equation}
and
\begin{equation}\label{biastheta}
B(\hat{\theta}_A) = \sum_{R,S,U} \kappa^{AR}\kappa^{SU} \left\{\kappa_{RS}^{(U)} - \frac{1}{2}\kappa_{RSU}\right\} - \frac{1}{2}\sum_{R,s,u} \kappa^{AR}\kappa^{su} \kappa_{Rsu}.
\end{equation}
These terms together with the cumulants needed to obtain them are given in the Appendix. After some tedious algebra, we
arrive at the following expression, in matrix form, for the second order bias of $\hat{\beta}$:
$$B(\hat{\beta}) = K^\beta \tilde{X}^T \Phi  M_1 Z_\beta  - \frac{1}{2} K^\beta \tilde{X}^T \Phi W_\beta E \mathbf{1},$$
where $K^\beta = K_\beta^{-1} = (\tilde{X}^T\Phi W_\beta \tilde{X})^{-1}$, $\mathbf{1}$ is an $n\times 1$ vector of ones, $Z_\beta$ is the $n\times 1$ dimensional vector
containing the diagonal elements of $\tilde{X}^T K^\beta \tilde{X}$, $W_\beta$ was defined in Section 2,
$E = {\rm diag}\left( {\rm tr}(\tilde{X}_i K^\beta)\right)$, $\tilde{X}_i$ is a $p\times p$ matrix with elements \label{formula} $\partial^2\eta_{1i}/\partial\beta_r\partial\beta_s$, and
\begin{equation}\label{m1}
M_1 = {\rm diag} \left(\frac{1}{2} \left\{(2d_{2i}'-d_{3i}) \left(\frac{d\mu_i}{d\eta_{1i}}\right)^3 + d_{2i} \frac{d\mu_i}{d\eta_{1i}}\frac{d^2\mu_i}{d\eta_{1i}^2}\right\} \right).
\end{equation}
Let $\omega_\beta = W_\beta^{-1}  M_1 Z_\beta$, thus the ${\cal O}(n^{-1})$ bias of $\hat{\beta}$ can be written
as
\begin{equation}\label{biasbetareg}
B(\hat{\beta}) = (\tilde{X}^T \Phi W_\beta \tilde{X})^{-1} \tilde{X}^T \Phi W_\beta (\omega_\beta - (1/2)E\mathbf{1}).
\end{equation}
Therefore, the ${\cal O}(n^{-1})$ bias of $\hat{\beta}$ (\ref{biasbetareg}) is easily obtained as the vector of regression
coefficients in the formal weighted linear regression of $\xi_\beta =  \omega_\beta - (1/2)E\mathbf{1}$ on the columns of $\tilde{X}$
with $\Phi W_\beta$ as weight matrix.

The ${\cal O}(n^{-1})$ bias (\ref{biasbetareg}) is expressed as the sum of two quantities: (i) $B_1 = (\tilde{X}^T \Phi W_\beta \tilde{X})^{-1} \tilde{X}^T \Phi W_\beta\omega_\beta$,
the bias for the MLE of the parameter $\beta$ on a linear dispersion regression with dispersion covariates with model matrix
$\tilde{X}$ and $\tilde{Z}$, and thus generalizes, for instance, the expressions obtained by Cordeiro and McCullagh (1991), and (ii) an additional quantity $B_2 = -(1/2) (\tilde{X}^T \Phi W_\beta \tilde{X})^{-1} \tilde{X}^T \Phi W_\beta E\mathbf{1}$
due to the nonlinearity of the function $f_1(x_i;\beta)$, and which vanishes if $f_1$ is linear with respect to $\beta$,
further, this expression generalizes, for instance, the expression obtained by Paula (1992).

Moving to the bias of $\hat{\theta}$, we have, after a tedious algebra on (\ref{biastheta}), the following expression for
the ${\cal O}(n^{-1})$ bias of $\hat{\theta}$:
$$B(\hat{\theta}) = K^\theta \tilde{Z}^T \{M_2 Z_\theta  - M_3 Z_\beta \} - \frac{1}{2} K^\theta \tilde{Z}^T W_\theta F\mathbf{1},$$
where $K^\phi = K_\phi^{-1} = (\tilde{Z}^T W_\theta \tilde{Z})^{-1}$, $Z_\theta$ is the $n\times 1$ dimensional vector
containing the diagonal elements of $\tilde{Z}^T K^\theta \tilde{Z}$, $W_\theta$ was defined in Section 2,
$F = {\rm diag}\left( {\rm tr}(\tilde{Z}_i K^\theta)\right)$, $\tilde{Z}_i$ is a $q\times q$ matrix with elements \label{formula2} $\partial^2\eta_{2i}/\partial\theta_R\theta_S$,
\begin{equation}\label{m2m3}
\begin{array}{c}
M_2 = {\rm diag}\left( \frac{1}{2}\left\{(2\alpha_{2i}'-\alpha_{3i}) \left(\frac{d\phi_i}{d\eta_{2i}}\right)^3 + \alpha_{2i}\frac{d\phi_i}{d\eta_{2i}}\frac{d^2\phi_i}{d\eta_{2i}^2}\right\}\right),\\
M_3 = {\rm diag}\left(\frac{d_{2i}}{2}\left(\frac{d\mu_i}{d\eta_{1i}}\right)^2\frac{d\phi_i}{d\eta_{2i}} \right).
\end{array}
\end{equation}
Let now, $\omega_\theta = W_\theta^{-1}\{ M_2 Z_\theta - M_3 Z_\beta \}$, then, we can express the ${\cal O}(n^{-1})$ bias
of $\hat{\theta}$ as
\begin{equation}\label{biasthetareg}
B(\hat{\theta}) = (\tilde{Z}^T W_\theta \tilde{Z})^{-1} \tilde{Z}^T W_\theta(\omega_\theta - (1/2) F \mathbf{1}).
\end{equation}

Thus, analogously to the ${\cal O}(n^{-1})$ bias of $\hat{\beta}$, the ${\cal O}(n^{-1})$ bias of $\hat{\theta}$ can
be obtained as the vector of regression coefficients in the formal weighted linear regression of $\xi_\theta = \omega_\theta - (1/2) F \mathbf{1}$
on the columns of $\tilde{Z}$ with $W_\theta$ as weight matrix.

Again, the ${\cal O}(n^{-1})$ bias (\ref{biasthetareg}) is expressed as the sum of two quantities: (i) $Q_1 =
(\tilde{Z}^T W_\theta \tilde{Z})^{-1} \tilde{Z}^T W_\theta\omega_\theta$, the bias of the parameter $\theta$ for a linear
dispersion regression with dispersion covariates with model matrices $\tilde{X}$ and $\tilde{Z}$, which generalizes the results obtained by
Botter and Cordeiro (1998), and (ii) $Q_2 =
-(1/2) (\tilde{Z}^T W_\theta \tilde{Z})^{-1} \tilde{Z}^T W_\theta F \mathbf{1}$ that is due to the nonlinearity of the functions
$f_1(x_i;\beta)$ and $f_2(z_i;\theta)$, and which vanishes if both $f_1$ and $f_2$ are linear in $\beta$ and $\theta$, respectively.

Now, let $B(\hat{\zeta}) = (B(\hat{\beta})^T,B(\hat{\theta})^T)^T$, we can then define our first bias-corrected estimator $\tilde{\zeta}$ as
$$\tilde{\zeta} = \hat{\zeta} - \hat{B}(\hat{\zeta}),$$
where $\hat{B}(\hat{\zeta})$ denotes the MLE of $B(\hat{\zeta})$, that is, the unknown parameters are replaced by their MLEs. Since the bias $B(\hat{\zeta})$
is of order ${\cal O}(n^{-1})$, it is not difficult to show that the asymptotic normality $\sqrt{n}\left(\tilde{\zeta}-\zeta\right)\stackrel{d}{\rightarrow}N_{p+q}(0,J^{-1}(\zeta))$
still holds, where, as before, we assume that $J(\zeta) = \lim_{n\to\infty} K(\zeta)/n$ exists and is nonsingular. From the asymptotic normality
of $\tilde{\zeta}$, we have that $\tilde{\zeta}_a \pm q_{1-\alpha/2}\left\{K(\tilde{\zeta})^{aa}\right\}^{1/2}$, for $a=1,\ldots,p,p+1,\ldots,p+q$.
The asymptotic variance of $\tilde{\zeta}_a$ is estimated by $K(\tilde{\zeta})^{aa}$, where $K(\tilde{\zeta})^{aa}$ is the $a$th diagonal element of
the inverse of the Fisher's information matrix evaluated at $\tilde{\zeta}$.

The last approach we consider here, to bias-correcting MLEs of the regression parameters is based upon the numerical estimation of the bias
through the bootstrap resampling scheme introduced by Efron (1979).  Let
${\bf y} = (y_1,\ldots, y_n)^{\top}$ be a random sample of size $n$, where each element is a random draw from the random
variable $Y$ which has the distribution function $F=F(\zeta)$. Here, $\zeta$ is the parameter that indexes the distribution, and
is viewed as a functional of $F$, i.e., $\zeta = t(F)$. Finally, let $\hat{\zeta}$ be an estimator of $\zeta$ based on ${\bf y}$; we write $\hat{\zeta} = s({\bf y})$.

The application of the bootstrap method consists in obtaining, from the original sample ${\bf y}$, a large number
of pseudo-samples ${\bf y}^{*} = (y_{1}^{*},\ldots, y_{n}^{*})^{\top}$, and then extracting information from these samples to improve
inference. Bootstrap methods can be classified into two classes, depending on how the sampling is performed: parametric and nonparametric.
In the parametric version, the bootstrap samples are obtained from $F(\hat{\zeta})$, which we shall denote as $F_{\hat{\zeta}}$, whereas in the nonparametric version they are
obtained from the empirical distribution function $\hat{F}$, through sampling with replacement. Note that the nonparametric bootstrap does not entail parametric assumptions.

Let $B_{F}(\hat{\zeta},\zeta)$ be the bias of the estimator $\hat{\zeta} = s({\bf y})$, that 
is, 
\[B_{F}(\hat{\zeta}, \zeta) = {\rm E}_{F}[\hat{\zeta} - \zeta] = {\rm E}_{F}[s({\bf y})] - t(F),\]
where the subscript $F$ indicates that expectation is taken with respect to $F$. The bootstrap estimators of
the bias in the parametric and nonparametric versions are obtained by replacing the true distribution $F$, which generated the original
sample,
with $F_{\hat{\zeta}}$ and $\hat{F}$, respectively, in the above expression.
Therefore, the parametric and nonparametric estimates of the bias are given, respectively, by
\[B_{F_{\hat{\zeta}}}(\hat{\zeta}, \zeta) = {\rm E}_{F_{\hat{\zeta}}}[s({\bf y})] - t(F_{\hat{\zeta}})
  \quad{\rm and}\quad B_{\hat{F}}(\hat{\zeta}, \zeta) = {\rm E}_{\hat{F}}[s({\bf y})] - t(\hat{F}).\]

\noindent If $B$ bootstrap samples $({\bf y}^{*1}, {\bf y}^{*2}, \ldots, {\bf y}^{*B})$ are generated independently from the original
sample ${\bf y}$, and the respective boostrap replications $(\hat{\zeta}^{*1}, \hat{\zeta}^{*2}, \ldots, \hat{\zeta}^{*B})$
are calculated,
where $\hat{\zeta}^{*b} = s({\bf y}^{*b})$, $b = 1, 2, \ldots, B$, then it is possible to approximate the bootstrap expectations 
${\rm E}_{F_{\hat{\zeta}}}[s({\bf y})]$ and ${\rm E}_{\hat{F}}[s({\bf y})]$
by the mean $\hat{\zeta}^{*(\cdot)} = \frac{1}{B}\sum_{b=1}^{B}\hat{\zeta}^{*b}$.
Therefore, the bootstrap bias estimates based on $B$ replications of $\hat{\zeta}$ are
\begin{equation}\label{biasboot}
\hat{B}_{F_{\hat{\zeta}}}(\hat{\zeta}, \zeta) = \hat{\zeta}^{*(\cdot)} - s({\bf y})\quad{\rm and}\quad
\hat{B}_{\hat{F}}(\hat{\zeta}, \zeta) = \hat{\zeta}^{*(\cdot)} - s({\bf y}),
\end{equation}

\noindent for the parametric and nonparametric versions, respectively.

By using the two bootstrap bias estimates presented above, we arrive at the following two bias-corrected, to order ${\cal O}(n^{-1})$, estimators:
\begin{eqnarray*}
\overline{\zeta}_{1} &=& s({\bf y}) - \hat{B}_{F_{\hat{\zeta}}}(\hat{\zeta}, \zeta) = 2\hat{\zeta} - \hat{\zeta}^{*(\cdot)},\\
\overline{\zeta}_{2} &=& s({\bf y}) - \hat{B}_{\hat{F}}(\hat{\zeta}, \zeta) = 2\hat{\zeta} - \hat{\zeta}^{*(\cdot)}.
\end{eqnarray*}
The corrected estimates $\overline{\zeta}_{1}$ and $\overline{\zeta}_{2}$ were called 
constant-bias-correcting (CBC) estimates by MacKinnon and Smith (1998).

Since we are dealing with regression models and not with a random sample we need some minor modifications to the algorithm given above.

For the nonparametric case, assume we want to fit a regression model with response variable $y$ and predictors $x_1,\ldots,x_{q_1},z_1,\ldots,z_{q_2}$.
We have a sample of $n$ observations $p_i^T = (y_i,x_{i1},\ldots,x_{iq_1},z_{i1},\ldots,z_{iq_2})$, $i=1,\ldots,n$. Thus we use the nonparametric bootstrap
method described above to obtain $B$ bootstrap samples of the $p_i^T$, fit the model and save the coefficients from each bootstrap sample. We
can then obtain bias corrected estimates for the regression coefficients using the methods described above. This is the so-called Random-$x$ resampling.

For the parametric case, assume we have the same model as for the nonparametric case, we thus obtain the estimates $\hat{\mu}_i$ and $\hat{\phi}_i$
(such as in our case where the distribution is indexed by $\mu$ and $\phi$) and using the
parametric method described above, we obtain $B$ bootstrap samples for $\hat{y}_i$ from the distribution $F(\hat{\mu}_i,\hat{\phi}_i)$, $i=1,\ldots,n$. We would then
regress each set of bootstrapped values $y_b^{\ast}$ on the covariates $x_1,\ldots,x_{q_1},z_1,\ldots,z_{q_2}$ to obtain bootstrap replications of the regression
coefficients. We can, again, obtain bias corrected estimates for the regression coefficients using the methods described above. This method is called
Fixed-$x$ resampling.

\section{Bias correction of the MLEs of $\mu$ and $\phi$}\label{biasmuphi}
In this Section we obtain the results that are the most valuable to the practioners, namely, the ${\cal O}(n^{-1})$ bias of $\mu$ and of $\phi$,
since, for practioners, the interest in a data analysis relies on sharp estimates of the responses and of the precision parameters. The fact that
these results must be computed apart comes from the fact that if $\ddot{\beta}$ and $\ddot{\theta}$ are bias-free estimators, to order ${\cal O}(n^{-1})$, it is not
true, in general, that $\ddot{\mu}_i = g_1^{-1}( f_1(x_i;\ddot{\beta}))$ and $\dot{\phi}_i = g_2^{-1}(f_2(z_i;\ddot{\theta}))$ will also be bias-free to
order ${\cal O}(n^{-1})$. Nevertheless, for practioners, it is even more important to correct the means of the responses and the precision parameters than
correcting the regression parameters. 

We shall first obtain the ${\cal O}(n^{-1})$ bias of the MLEs of $\eta_1$ and $\eta_2$. Using (\ref{regr}) we find, by Taylor expansion, that to order ${\cal O}(n^{-1})$:
$$f_1(x_i^T; \hat{\beta}) - f_1(x_i^T; \beta) = \nabla_\beta(\eta_{1i})^T(\hat{\beta}-\beta) + \frac{1}{2}(\hat{\beta}-\beta)^T \tilde{X}_i (\hat{\beta}-\beta),$$
and
$$f_2(z_i^T; \hat{\theta}) - f_2(z_i^T; \theta) = \nabla_\theta(\eta_{2i})^T (\hat{\theta}-\theta) + \frac{1}{2}(\hat{\theta}-\theta)^T\tilde{Z}_i (\hat{\theta}-\theta),$$
where $\nabla_\beta(\eta_{1i})$ is a $p\times 1$ vector with the derivatives $\partial\eta_{1i}/\partial\beta_r$, $\nabla_\theta(\eta_{2i})$ is a
$q\times 1$ vector with the derivatives $\partial\eta_{2i}/\partial\theta_R$.

Thus, taking expectations on both sides of the above expression yields to this order
$$B(\hat{\eta_1}) = \tilde{X}B(\hat{\beta}) + \frac{1}{2}E,$$
and
$$B(\hat{\eta_2}) = \tilde{Z}B(\hat{\theta}) + \frac{1}{2}F,$$
where, $E$ and $F$ were defined in Section 3, and we used the fact that $K^{\beta}$ and $K^{\theta}$ are the asymptotic covariance matrices of
$\hat{\beta}$ and $\hat{\theta}$, respectively.

>From similar calculations we obtain to order ${\cal O}(n^{-1})$
$$B(\hat{\mu}_i) = B(\hat{\eta}_{1i})\frac{d\mu_i}{d\eta_{1i}} + \frac{1}{2}{\rm Var}(\hat{\eta_{1i}})\frac{d^2\mu_i}{d\eta_{1i}^2}$$
and
$$B(\hat{\phi}_i) = B(\hat{\eta}_{2i})\frac{d\phi_i}{d\eta_{2i}} + \frac{1}{2}{\rm Var}(\hat{\eta_{2i}})\frac{d^2\mu_i}{d\eta_{2i}^2}.$$

Let $T_1$ and $T_2$ be as in Section 2, further, let $S_1 = {\rm diag}(d^2\mu_i/d\eta_{1i}^2)$ and $S_2 = {\rm diag}(d^2\phi_i/d\eta_{2i}^2)$.
Then, we can write the above expressions in matrix notation as
\begin{equation}\label{biasmu}
B(\hat{\mu}) = \frac{1}{2}T_1(2\tilde{X}B(\hat{\beta})+E) + \frac{1}{2} S_1 Z_{\beta}
\end{equation}
and
\begin{equation}\label{biasphi}
B(\hat{\phi}) = \frac{1}{2}T_2(2\tilde{Z}B(\hat{\theta}) + F) + \frac{1}{2} S_2 Z_{\theta},
\end{equation}
where $Z_{\beta}$ and $Z_{\theta}$ were defined in Section 3, and the asymptotic covariance matrices of $\hat{\eta_1}$ and
$\hat{\eta_2}$ are $\tilde{X}K^{\beta}\tilde{X}^T$ and $\tilde{Z}K^{\theta}\tilde{Z}^T$, respectively.

If we combine (\ref{biasmu}) and (\ref{biasphi}) with (\ref{biasbetareg}) and (\ref{biasthetareg}), we will have the following explicit expressions for the
${\cal O}(n^{-1})$ biases of $\hat{\mu}$ and $\hat{\phi}$, respectively:
$$B_1(\hat{\mu}) = \frac{1}{2}T_1(2\tilde{X}K^{\beta}\tilde{X}^T\Phi W_\beta (\omega_\beta - (1/2)E\mathbf{1})+E) + \frac{1}{2}S_1Z_{\beta}$$
and
$$B_1(\hat{\phi})= \frac{1}{2}T_2(2\tilde{Z}K^{\theta}\tilde{Z}^T W_\theta(\omega_\theta - (1/2) F \mathbf{1})+F) + \frac{1}{2}S_2Z_{\theta}.$$

Lastly, we can use the bootstrap-based ${\cal O}(n^{-1})$ biases to define, bias corrected estimators of $\hat{\mu}$ and $\hat{\phi}$ to this order.
Then, let $\hat{B}_{F_{\hat{\zeta}}}(\hat{\beta})$ be the vector formed by the first $p$ elements of the vector $\hat{B}_{F_{\hat{\zeta}}}(\hat{\zeta}, \zeta)$
defined in equation (\ref{biasboot}), $\hat{B}_{F_{\hat{\zeta}}}(\hat{\theta})$ be the vector formed by the last $q$ elements of the vector $\hat{B}_{F_{\hat{\zeta}}}(\hat{\zeta}, \zeta)$,
and define $\hat{B}_{\hat{F}}(\hat{\beta})$ and $\hat{B}_{\hat{F}}(\hat{\theta})$ analogously from the vector $\hat{B}_{\hat{F}}(\hat{\zeta}, \zeta)$ also in equation (\ref{biasboot}).
Thus, we have the following alternative expressions for the ${\cal O}(n^{-1})$ biases of $\hat{\mu}$ and $\hat{\phi}$, respectively:
$$B_2(\hat{\mu}) = \frac{1}{2}T_1(2\tilde{X}\hat{B}_{F_{\hat{\zeta}}}(\hat{\beta})+F) + \frac{1}{2} S_1 P_{\beta\beta}\hbox{~and~}B_3(\hat{\mu}) = \frac{1}{2}T_1 (2\tilde{X}\hat{B}_{\hat{F}}(\hat{\beta})+F) + \frac{1}{2} S_1 P_{\beta\beta},$$
and
$$B_2(\hat{\phi}) = \frac{1}{2}T_2(2\tilde{Z}\hat{B}_{F_{\hat{\zeta}}}(\hat{\theta}) + G) + \frac{1}{2} S_2 P_{\theta\theta} \hbox{~and~}B_3(\hat{\phi}) = \frac{1}{2}T_2 (2\tilde{Z}\hat{B}_{\hat{F}}(\hat{\theta})+G) + \frac{1}{2} S_2 P_{\theta\theta}.$$

Therefore, we are now able to define the following second-order bias-corrected estimators for $\hat{\mu}$ and $\hat{\phi}$:
$$\tilde{\mu} = \hat{\mu} - \hat{B}_1(\hat{\mu}),\quad\overline{\mu}_1 = \hat{\mu}-\hat{B}_2(\hat{\mu})\hbox{~~and~~} \overline{\mu}_2 = \hat{\mu}-\hat{B}_3(\hat{\mu})$$
and
$$\tilde{\phi} = \hat{\phi} - \hat{B}_1(\hat{\phi}),\quad \overline{\phi}_1 = \hat{\phi}-\hat{B}_2(\hat{\phi})\hbox{~~and~~} \overline{\phi}_2 = \hat{\phi}-\hat{B}_3(\hat{\phi}),$$
where, for $j=1,2$ and $3$, $\hat{B}_j(\cdot)$ denotes the MLE of $B_j(\cdot)$, that is, the unknown parameters are replaced by their MLEs.

\section{Some special cases}
In this section we examine some special cases of the formula . Some other important cases could also be easily obtained
because of the advantage of this formula that involves only simple operations on suitably defined matrices and can be easily implemented
in statistical packages or in a computer algebra system such as Mathematica or Maple.

Table \ref{tabelalink} below shows the most common link functions and the quantities needed in order to compute the biases of the MLEs of the parameters $\beta$ and $\theta$. In Table \ref{tabelalink}: $\Phi(\cdot)$ denotes
the standard normal distribution function; $f(x) = 1/\sqrt{2\pi}\exp\{-1/2 x^2\}$ is the density of a
standard normal distribution; and $f'(x) = -x/\sqrt{2\pi} \exp\{-1/2 x^2\}$ is the derivative of the density
of a standard normal distribution.

\begin{table}[hbt]
\caption{Most common link functions.}
\begin{center}
\begin{tabular}{lccc}

\hline
Link & Formula & ${d\mu}/{d\eta}$&${d^2\mu}/{d\eta^2}$ \\
   
\hline
Logit            &$\log\left({\mu}/{(1-\mu)}\right)=\eta$&$\mu(1-\mu)$&$\mu(1-\mu)(1-2\mu)$ \\
Probit           &$\Phi^{-1}(\mu)=\eta$&$f(\Phi^{-1}(\mu))$&$f'(\Phi^{-1}(\mu))$\\
Log              &$\log(\mu)=\eta$&$\mu$&$\mu$\\
Identity         &$\mu=\eta$&$1$&$0$\\
Reciprocal       &$\mu^{-1}=\eta$&$-\mu^2$&$2\mu^3$\\
Square reciprocal&$\mu^{-2}=\eta$&${-\mu^3}/{2}$&${3\mu^5}/{4}$\\
Square Root      &$\sqrt{\mu}=\eta$&$2\sqrt{\mu}$&$2$\\
C-loglog         &$\log(-\log(1-\mu))=\eta$&$-\log(1-\mu)(1-\mu)$&$-(1-\mu)\log(1-\mu)\times$\\
                 &                         &                     &$\times(1+\log(1-\mu))$\\
Tangent          &$\tan(\mu)=\eta$&$\cos(\mu)^2$&$2\cos(\mu)^3\sin(\mu)$\\
\hline
\end{tabular}
\end{center}
\label{tabelalink}
\end{table}
\subsection{Generalized linear models with dispersion covariates}
The results obtained in this subsection generalize the results obtained in the articles by Cordeiro and McCullagh (1991) and Botter and Cordeiro (1998).

We begin by analysing the ${\cal O}(n^{-1})$ bias of the parameter $\beta$. Here, the function $t(\cdot,\cdot)$ has the form $t(y,\theta) = y\theta - b(\theta),$ where $b'(\theta) = \mu$. Thus, consider the function
$\tau(\theta) = b'(\theta)$, $\tau(\cdot)$ is called the mean value mapping, the variance function is related to the mean value mapping by
$d\tau^{-1}(\mu)/d\mu = V(\mu)^{-1}$. We have that $t\{y,\tau^{-1}(\mu)\} = y\tau^{-1}(\mu) - b\{\tau^{-1}(\mu)\}$.
For generalized linear models, $d_2 = - V^{-1}$ and $d_3 = 2V^{-2}V^{(1)}$, where $V^{(1)} = dV(\mu)/d\mu$. Thus,
the matrix $W$ reduces to $W = \{V^{-1}(d\mu/d\eta)^2\}$. The local model matrix $\widetilde{X}$ also reduces to the matrix $X$ from $h(\mu_i) = \eta_i = x_i^T \beta$ and $E$ vanishes. Further, we have that
$$M_1 = \diag\left\{-\frac{1}{2} V^{-1} \frac{d\mu}{d\eta}\frac{d^2\mu}{d\eta^2}\right\},$$
which is precisely the result obtained by Cordeiro and McCullagh (1991).\\

Table \ref{tabelaexpf} shows the distributions in the exponential family, along with the quantities needed to obtain the bias.

\begin{table}[!hb]
\caption{Exponential Family}
\begin{center}
\begin{tabular}{lccc}
\hline
Distribution&$V$&$V^{(1)}$&$V^{(2)}$\\
\hline
Normal      &$1$&$0$&$0$\\
Poisson     &$\mu$&$1$&$0$\\
Binomial    &$\mu(1-\mu)$&$1-2\mu$&$-2$\\
Gamma       &$\mu^2$&$2\mu$&$2$\\
Inv. Gauss.&$\mu^3$&$3\mu^2$&$6\mu$\\
\hline
\end{tabular} 
\end{center}
\label{tabelaexpf}
\end{table}

We now move to the bias for the dispersion parameter $\theta$. So, let's consider the two-parameter full exponential family distributions with canonical parameters $\phi$ and $\phi\vartheta$. Therefore, we have $a(\phi,y) = \phi c(y) + a_1(\phi) + a_2(y)$, where $c(\cdot)$ is a known appropriate function. Then
it turns out that $\alpha_2 = a_1''(\phi)$ and $\alpha_3 = \alpha_2'= a_1'''(\phi)$. Then, using (\ref{m2m3}), we have that
$$M_2 = \diag\left\{\frac{1}{2} \left[a_1'''(\phi) \left(\frac{d\phi}{d\eta_2}\right)^3 +a_1''(\phi)\frac{d\phi}{d\eta_2}\frac{d^2\phi}{d\eta_2^2} \right]\right\},$$
and
$$M_3 = \diag\left\{-\frac{1}{2}V^{-1}\left(\frac{d\mu}{d\eta_1}\right)^2\frac{d\phi}{d\eta_2} \right\}.$$

The expressions above agrees with the formula presented by Botter and Cordeiro (1998).

Table \ref{expfamdisp} presents the values of the derivatives of the function $a_1$ for the distributions in the exponential family. In Table \ref{expfamdisp}, $\psi^{(m)}(\cdot)$, $m=0,1,\ldots,$ is the polygamma function defined by $\psi^{(m)}(x) = \left( d^{m+1}/dx^{m+1}\right)\log\Gamma(x),x>0.$

\begin{table}[!hb]
\caption{Exponential Family}
\begin{center}
\begin{tabular}{lccc}
\hline
Distribution&$a_1(\phi)$&$a_1''(\phi)$&$a_1'''(\phi)$\\
\hline
Normal      &$\log\sqrt{\phi}$&$-\frac{1}{2\phi^2}$&$\frac{1}{\phi^3}$\\
Gamma       &$\phi\log(\phi)-\log\Gamma{\phi}$&$\frac{1}{\phi} + \psi'(\phi)$&$-\frac{1}{\phi^2} + \psi''(\phi)$\\
Inv. Gauss.&$\log\sqrt{\phi}$&$-\frac{1}{2\phi^2}$&$\frac{1}{\phi^3}$\\
\hline
\end{tabular} 
\end{center}
\label{expfamdisp}
\end{table}
\subsection{Exponential family nonlinear models with dispersion covariates}
This model generalizes the generalized linear model with dispersion covariates. Recently Simas and Cordeiro (2009) provided ajusted Pearson residuals for exponential family nonlinear models. We only have the ${\cal O}(n^{-1})$ bias computed in the literature for the exponential family nonlinear models with constant dispersion parameter (see Paula, 1991). The results for the exponential family nonlinear model with disperion covariates are new. 

Let us consider the same parameterization from above, i.e., $t\{y,\tau^{-1}(\mu)\} = y\tau^{-1}(\mu) - b\{\tau^{-1}(\mu)\}$, with
$d\tau^{-1}(\mu)/d\mu = V(\mu)^{-1}$. Then, the matrices $M_1, M_2$ and $M_3$ are the same as the ones computed in the previous subsection 

We now present in Table \ref{tabelaexpdisp} the results for two distributions that belong to the class of \emph{exponential dispersion models} introduced by J\o rgensen (1987). 

\begin{table}[!h]
\caption{Exponential dispersion models.}
\begin{center}
\begin{tabular}{lccc}
\hline
Distribution & $d_2$&$d_2'$&$d_3$\\   
\hline
GHS                      &$\frac{2}{(\mu^2+1)^2}$ &$\frac{-8\mu}{(\mu^2+1)^3}$& $-\frac{(2\mu^3+10\mu)}{(\mu^2+1)^3}$\\
Neg. Bin.        &$\frac{1}{\mu}-\frac{1}{1-\mu}$&$-\left[\frac{1}{\mu^2}-\frac{1}{(1-\mu)^2}\right]$&$\frac{2}{(1+\mu)^2}-\frac{2}{\mu^2}$\\
Power Var.           &$-\mu^{-p}$&$p\mu^{-(p+1)}$&$2p\mu^{-(p+1)}$\\
Exp. Var.     &$-e^{-\beta\mu}$&$\beta e^{-\beta\mu}$&$2\beta e^{-\beta\mu}$\\
\hline
\end{tabular} 
\end{center}
\label{tabelaexpdisp}
\end{table}

Among these distributions are the generalized hyperbolic secant and the negative binomial. Our results
can be applied for a very rich class of models discussed in detail in J\o rgensen's (1997b) book. He presented several exponential dispersion models in \eqref{dens} including the Tweedie class of distributions with power variance function defined by taking $V(\mu)=\mu^{\delta}$ and the cumulant generator function
$b_{\delta}(\theta)$ for $\delta\neq1,2$ by
$$b_{\delta}(\theta) = (2-\delta)^{-1} \left\{(1-\delta)\theta \right\}^{\frac {\delta-2}
{\delta-1}},$$
and $b_{1}(\theta)=\exp(\theta)$ and $b_{2}(\theta)=-\log(-\theta)$. 
We recognize for $\delta=0,2$ and $3$, the cumulant generator corresponding 
to the normal, gamma and inverse Gaussian distributions, respectively. 
There exist continuous exponential dispersion models generated by extreme stable distributions with 
support $\mathbb{R}$ and positive stable distributions, respectively, 
when $\delta \leq 0$ and $\delta \geq 2$ and compound Poisson distributions 
for $1<\delta<2$. We also would like to remark that there exists an exponential dispersion model with exponential
variance function, $V(\mu) = e^\mu$, for more details see the book of Jorgensen (1997b).

Finally, it is noteworthy that this special case has not been treated in the literature until now.
\subsection{Proper dispersion models with dispersion covariates}
For proper dispersion models, the formula (\ref{m1}) have no reduction, since the only difference of a proper dispersion
model from a dispersion model is the form of the function $a(\cdot,\cdot)$ which can be decomposed into $a(\phi,y) = a_1(\phi) + a_2(y)$. We will now give the expression for the matrices $M_2$ and $M_3$. First we note that for this case $\alpha_2 = a_1''(\phi)$ and $\alpha_3 = \alpha_2'= a_1'''(\phi)$. Then, using (\ref{m2m3}), we have that
$$M_2 = \diag\left\{\frac{1}{2} \left[a_1'''(\phi) \left(\frac{d\phi}{d\eta_2}\right)^3 +a_1''(\phi)\frac{d\phi}{d\eta_2}\frac{d^2\phi}{d\eta_2^2} \right]\right\},$$
and
$$M_3 = \diag\left\{-\frac{1}{2}V^{-1}\left(\frac{d\mu}{d\eta_1}\right)^2\frac{d\phi}{d\eta_2} \right\}.$$
Note that even though the form of $a(\phi,y)$ for this case is different from the form of $a(\phi,y)$ for the two-parameter full exponential family model, the expressions for $M_2$ and $M_3$ are equal.

But to illustrate the idea on a particular example of proper dispersion model, we will consider the \emph{von Mises regression model}. Then, we now move to von Mises regression models which are quite useful for modelling circular data; see Fisher (1993) and Mardia (1972).
Here,the density is given by
\begin{equation}\label{vonmises}
\pi(y;\mu,\phi) = \frac{1}{2\pi I_0(\phi)} \exp\{\phi\cos(y-\mu)\},
\end{equation}
where, $-\pi<y\leq \pi$, $-\pi<\mu\leq \pi$, $\phi>0$, and $I_v$ denotes the modified Bessel function of the first kind and order $v$
(see Abramowitz and Stegun, 1970, Eq. 9.6.1). The density in (\ref{vonmises}) is symmetric around $y=\mu$ which is the mode and
the circular mean of the distribution. $\phi$ is a precision parameter in the sense that the larger the value of $\phi$ the more
concentrated the density around $\mu$ gets. It is clear that the density (\ref{vonmises}) is a proper dispersion model, since
$t(y,\mu) = \cos(y-\mu)$ and $a_1(\phi) = \log\{I_0(\phi)\}$. We now begin by investigating the skewness for the parameters $\beta$. Then, it is possible to show that $E\{\sin(Y-\mu)\} = 0$ and
$E[\{\cos(Y-\mu)\}^2] = 1-\phi^{-1}r(\phi)$, where $r(\phi) = I_1(\phi)/I_0(\phi)$, these results yield $d_2 = -r(\phi),$
$d_3 = 0$ and $d_2' = 0$. Further, we have that the matrix $W = {\rm diag}\{(d\mu/d\eta)^2 r(\phi)\}$.

Note initially that $I_0'(\phi) = I_1(\phi)$ and $I_1'(\phi) = I_0(\phi) - I_1(\phi)/\phi$ (Abramowitz and Stegun, 1970; equations 9.6.26 and 9.6.27). Then, $a_1''(\phi) = r'(\phi)$  and $a_1'''(\phi) = r''(\phi)$, where, as above, $r(\phi) = I_1(\phi)/I_0(\phi)$.

We have that 
$$M_1 = \diag\left\{-\frac{r(\phi)}{2} \frac{d\mu}{d\eta_1} \frac{d^2\mu}{d\eta_1^2} \right\},$$
$$M_2 = \diag\left\{\frac{r''(\phi)}{2}\left(\frac{d\phi}{d\eta_2}\right)^3 + \frac{r'(\phi)}{2}\frac{d\phi}{d\eta_2}\frac{d^2\phi}{d\eta_2^2}  \right\},$$
and
$$M_3 = \diag\left\{-\frac{r(\phi)}{2} \left(\frac{d\mu}{d\eta_1}\right)^2\frac{d\phi}{d\eta_2}\right\}.$$ 
We provide in Tables \ref{tabelaproper} the quantities needed for several distributions in the class of proper dispersion models.

\begin{table}[!t]
\caption{Proper Dispersion Models}
\begin{center}
\begin{tabular}{lccc}
\hline
Distribution & $d_2$&$d_2'$&$d_3$\\   
\hline
Rec. Gamma         &$-\mu^{-2}$&$2\mu^{-3}$&$2\mu^{-3}$\\
Log-Gamma                &$-1$&$0$&$1$\\
Rec. Inv. Gauss. &$-\mu^{-1}$&$\mu^{-2}$&$0$\\
Von-Mises                &$-r(\phi)$&$0$&$0$\\
\hline
\end{tabular} 
\end{center}
\label{tabelaproper}
\end{table}

In Table \ref{tabelaproper2} we give the derivatives of the function $a_1$ for several distributions in the class of proper dispersion models.

\begin{table}[!t]
\caption{Proper Dispersion Models}
\begin{center}
\begin{tabular}{lccc}
\hline
Distribution & $a_1(\phi)$&$a_1''(\phi)$&$a_1'''(\phi)$\\   
\hline
Rec. Gamma         &$\phi\log(\phi)-\log\Gamma{\phi}$&$\frac{1}{\phi} + \psi'(\phi)$&$-\frac{1}{\phi^2} + \psi''(\phi)$\\
Log-Gamma                &$\phi\log(\phi)-\log\Gamma{\phi}$&$\frac{1}{\phi} + \psi'(\phi)$&$-\frac{1}{\phi^2} + \psi''(\phi)$\\
Rec. Inv. Gauss. &$\log\sqrt{\phi}$&$-\frac{1}{2\phi^2}$&$\frac{1}{\phi^3}$\\
Von-Mises                &$\log I_0(\phi)$&$r'(\phi)$&$r''(\phi)$\\
\hline
\end{tabular} 
\end{center}
\label{tabelaproper2}
\end{table}

\subsection{Some other special cases}
We now investigate some special cases which were first studied by Cordeiro (1983). If we take $t(y,\theta) = y\mu-b(\mu)$,
(\ref{dens}) is a one parameter exponential family indexed by the canonical parameter $\mu$. Now, if in (\ref{dens}) we assume that $t(y,\mu)$ involves a known constant parameter $c$ for all observations, $t(y,\mu) = t(y,\mu,c)$ say, and that $\phi = 1$ and $a(\phi,y) = a(c,y)$. For doing this this, several models can be defined within the present framework: normal distribution $N(\mu,c^2\mu^2)$, lognormal $LN(\mu,c^2\mu^2)$, inverse Gaussian distribution $IG(\mu,c^2\mu^2)$ with mean $\mu$ and known constant coefficient of variation $c$, Weibull distribution $W(\mu,c)$ with mean $\mu$ and known constant shape parameter $c$. Here the normal and inverse Gaussian distribtuions are not standard generalized linear models since we are considering a different parameterization.

For these models, we have that $d_2 = -k_2 \mu^{-2}$, $d_3 = k_3 \mu^{-3}$ and $d_2' = 2k_2\mu^{-3}$, where $k_2$ and $k_3$ are known positive functions of $c$ (see Table \ref{tabelaspecial}). Then, we have the matrix $W = {\rm diag}\{k_2\mu^{-2}(d\mu/d\eta)^2\}$, and hence we are able to obtain the inverse of the information matrix, and the matrix $M_1$. Moreover, $w = k_2\mu^{-2}(d\mu/d\eta)^2$, and
$$M_1 = \diag\left\{ \frac{1}{2}\left[(4k_2-k_3)\mu^{-3}\left(\frac{d\mu}{d\eta_1} \right)^3 -k_2\mu^{-2} \frac{d\mu}{d\eta_1}\frac{d^2\mu}{d\eta_2^2} \right]\right\}. $$
\begin{table}[htb]
\caption{Values of $k_2$ and $k_3$ for the normal, inverse Gaussian, lognormal and Weibull distributions.}
\begin{center}
\begin{tabular}{ccc}
\hline
Model            &$k_2(c)$      & $k_3(c)$\\
\hline                           
Normal ($N(\mu,c^2\mu^2)$)   &$c^{-2}(1+2c^2)$         & $c^{-2}(6+10c^2)$ \\
Inverse Gaussian ($IG(\mu,c^2\mu^2)$) &$1/2 c^{-2}(1+c^2)$ & $c^{-2}(3+c^2)$  \\
Lognormal ($LN(\mu, c^2\mu^2)$)& $[\log(1+c^2)]^{-1}$ & $3[\log(1+c^2)]^{-1}$ \\
Weibull ($W(\mu,c)$) & $c^2$ & $c^2(c+3)$ \\
\hline
\end{tabular}
\end{center}
\label{tabelaspecial}
\end{table}
\section{Numerical Results}
In this section we present the results of some Monte Carlo simulation experiments,
where we study the finite-sample distributions of the MLEs of $\beta$ and $\theta$ along with their corrected versions proposed in this paper. We use a reciprocal gamma model with square root link and a log link in a nonlinear model for the dispersion
parameter
$$\sqrt{\mu_i} = \beta_0 + \beta_1 x_{1,i} + x_{2,i}^{\beta_2},$$
$${\rm log}\phi_i = \theta_0 + \theta_1 x_{1,i} + x_{2,i}^{\theta_2},\quad i=1,\ldots,n,$$ 
where the true values of the parameters were taken as $\beta_0 =
1/2$, $\beta_1 = 1$, $\beta_2=2$ and $\theta_0 = 1, \theta_1 = 2$ and $\theta_2 = 3$. Note also that here
the elements of the $n\times 3$ matrix $\tilde{X}$ are: $\tilde{X}(\beta)_{i,1} = 1; \tilde{X}(\beta)_{i,2} = x_{1,i}$, and
$\tilde{X}(\beta)_{i,3} = \log(x_{2,i}) x_{2,i}^{\beta_2}$. The explanatory
variables $x_1$ and $x_2$ were generated from the uniform U$(0,1)$
distribution for sample size $n=20$, and their values were held constant
throughout the simulations. The number of Monte Carlo replications
was set at $5,000$ and all simulations were performed using the
statistical software {\tt R}.

In each of the $5,000$ replications, we fitted the model and computed the MLEs $\hat{\beta}$,
$\hat{\theta}$, its corrected versions from the corrective method (Cox and Snell, 1968),
preventive method (Firth, 1993) and the bootstrap method both of its parametric and nonparametric versions (Efron, 1979).
The number of bootstrap replications was set to 500 for both bootstrap methods.

In order to analyze the results we computed, for each sample size and for each estimator, the mean of estimates, bias, variance and mean square error (MSE). Table \ref{resulsimul} present simulation results.

Lastly, in each replication we estimated the confidence interval for each parameter for each estimator, and verified if the true value of the parameter belonged to this estimated confidence interval. After that we obtained the average of the number of confidence intervals that contained the true parameter. In this way we were able to check if the estimated confidence interval was close to its nominal level of confidence. The confidence intervals were constructed following the strategies stated at the end of Section 2 and at Section 3.

Table \ref{resulsimul} presents simulation results for sample size $n=20$ with respect to the parameters $\beta$ and $\theta$. We begin by looking at the estimated biases, in absolute value, of the estimators. Initially, we note that for all parameters the biases of the corrective estimators were smaller than those of the original MLEs. However, for all parameters the biases of the preventive estimators were larger than those of the original MLEs. Moreover, not only the biases were larger but also the MSEs were larger as well, which shows that the preventive method does not work well for this model. The same phenomenon occurred in Ospina et al. (2006), which corroborates the idea that this method has some problems in beta regression models. We now observe that the MSE of the corrective estimators were smaller than those of the MLEs for all parameters, showing that the correction is effective. Moving to the bootstrap corrected-estimators, we note that the parametric bootstrap had the smallest MSE for all parameters, even though the biases were not the smallest. However, the MSEs were very close to the MSE of the corrective method, and the computation of the parametric bootstrap biases is computer intensive, whereas the corrective method is not. Lastly, we observe that for all parameters $\theta$ the MSE of the nonparametric bootstrap corrected estimators were smaller than those of the MLEs. Moreover, for the parameters $\beta$, the MSE of the nonparametric bootstrap corrected estimators were very close to those of the MLEs, showing that this method is satisfactory, and is very easy to implement by practitioners since no parametric assumptions are made. Therefore, for the small sample size $n=20$, we were able to conclude that the corrective method by Cox and Snell (1969) was successfully applied, as well as the bootstrap corrections.

\begin{table}[h!]
\caption{Simulation results.}
\begin{center}
\begin{tabular}{lcccc}
\hline
Parameter & MLE& Cox-Snell  & p-boot & np-boot\\
\hline
$\beta_0$  &0.6356 &0.5552  &0.5728 &0.6001 \\
Bias       &0.1356 &0.0552  &0.0728 &0.1001\\ 
Variance   &0.0716 &0.0707  &0.0683 &0.0755\\
MSE        &0.0899 &0.0737  &0.0735 &0.0855\\\\
$\beta_1$  &0.9383 &1.0220  &0.9535 &1.0519         \\
Bias       &-0.0617&0.0220  &-0.0465&0.0519\\
Variance   &0.0251 &0.0224  &0.0203 &0.0261\\
MSE        &0.0289 &0.0228  &0.0224 &0.0287\\\\
$\beta_2$  &1.8853 &2.0075  &1.9783 &1.9099\\
Bias       &-0.1147&0.0075  &-0.0217&-0.0901\\
Variance   &0.0348 &0.0316  &0.0289 &0.0331\\
MSE        &0.0479 &0.0317  &0.0293 &0.0412\\\\
$\theta_0$ &1.0531 &1.0211  &1.0248 & 1.0612\\
Bias       &0.0531 &0.0211  &0.0248 & 0.0612\\
Variance   &0.5805 &0.5332  &0.4669 &0.4841\\
MSE        &0.5833 &0.5336  &0.4675 &0.4878\\\\
$\theta_1$ &2.1077 &1.9934  &1.9872 &2.1067\\
Bias       &0.1077 &-0.0066 &-0.0128&0.1067\\
Variance   &0.3001 &0.2222  &0.2345 &0.2300\\
MSE        &0.3117 &0.2222  &0.2347 &0.2414\\\\
$\theta_2$ &3.0464 &3.0077  &3.0115 &3.0519\\
Bias       &0.0464 &0.0077  &0.0115 &0.0519\\
Variance   &0.1101 &0.0858  &0.0686 &0.0525\\
MSE        &0.1122 &0.0858  &0.0687 &0.0551\\
\hline
\end{tabular}
\end{center}
\label{resulsimul}
\end{table}
Table \ref{tabelan20} presents the simulation results for sample size $n=20$ with respect to coverage of the interval estimates on different nominal converages $1-\alpha=90\%, 95\%$ and $99\%$. All confidence intervals
were defined such that the probability that the true parameter value belongs to the interval is $1-\alpha$, the probability
that the true parameter value is smaller than the lower limit of the interval is $\alpha/2$ and the probability that the
value of the parameter is greater than the upper limit of the interval is $\alpha/2$ for $0<\alpha<1/2$.

We begin by noting that the confidence intervals induced by the Firth estimates had the worst coverage, and therefore are not reliable. Further, the MLE and the non-parametric bootstrap had a similar behavior. The best coverage is from the corrective method Cox-Snell, all the coverage were closer to the nominal level with the Cox-Snell than any other estimator. Finally the parametric bootstrap had a poor perfomance with respect to the coverage of the confidence interval. The reason for that, we believe, is that the bootstrap estimator had the smallest MSE, which was in fact, due to the fact that it had the smallest variance among all the other estimators as seen in Table \ref{resulsimul}, therefore the confidence intervals induced by the parametric bootstrap estimator had the smallest average length, which yielded this poor coverage.

\begin{table}[h!]
\caption{Coverage of the interval estimates of the parameters.}
\begin{center}
\begin{tabular}{clccccccccc}
\hline
$\alpha$& Estimator & {$\beta_0$}&{$\beta_1$} & {$\beta_2$}& {$\theta_0$}&{$\theta_1$} & {$\theta_2$}\\
    \hline
      &MLE      &0.8275&0.7928 &0.8079 &0.7911 &0.7245 &0.7811\\
      &Cox-Snell&0.8788&0.8213 &0.8710 &0.8482 &0.7826 &0.8296\\
$10\%$&p-boot   &0.8131&0.7897 &0.7771 &0.7642 &0.7155 &0.7721\\
      &np-boot  &0.8352&0.8013 &0.8239 &0.7965 &0.7307 &0.8004\\
\\                                                                  
      &MLE      &0.8827&0.8417 &0.8913 &0.8608 &0.8424 &0.8549\\
      &Cox-Snell&0.9279&0.8981 &0.9311 &0.9035 &0.8745 &0.8894\\
 $5\%$&p-boot   &0.8560&0.8100 &0.8475 &0.8382 &0.8133 &0.8351\\
      &np-boot  &0.8846&0.8592 &0.9022 &0.8768 &0.8488 &0.8673\\
\\                                                                  
      &MLE      &0.9592&0.9216 &0.9665 &0.9439 &0.9133 &0.9288\\
      &Cox-Snell&0.9771&0.9653 &0.9803 &0.9608 &0.9409 &0.9573\\
 $1\%$&p-boot   &0.9385&0.9037 &0.9318 &0.9194 &0.8867 &0.8999\\
      &np-boot  &0.9678&0.9284 &0.9621 &0.9518 &0.9175 &0.9334\\
\hline
\end{tabular}
\end{center}
\label{tabelan20}
\end{table}

Finally, we would like to remark that one may build hypothesis tests upon confidence intervals. Further, if the confidence level of the confidence interval is 1-$\alpha$, then the test based on this confidence interval will have significance level $\alpha$. Moreover, the tests based on the confidence intervals used in this article are equivalent to Wald tests. Therefore, the hypothesis tests based on the confidence intervals would have significance levels closest to the nominal level when using the corrective method.

\section{Conclusion}
We defined a general dispersion model which allows a regression structure on the
precision parameter, in such a way that the regression structures on both the mean and the precision
parameters are allowed to be nonlinear. Then, using the approximation theory developed
by Cox and Snell (1968), we
calculate the ${\cal O}\left (n^{-1}\right )$ bias for the MLEs for $\beta$ and $\theta$. 

The dispersion models extends the well-known generalized linear models and also the exponential family nonlinear models. It is also important to say that is also generalizes the class of Proper dispersion models introduced by J\o rgensen (1997a). Several properties and applications of dispersion models can be found on the excellent book of J\o rgensen (1997b).

Our results, thus, generalize, for instance, the formulae obtained by Cordeiro and McCullagh (1991), Paula (1992), Cordeiro and Vasconcellos (1999) and Botter and Cordeiro (1998). We then defined bias-free estimators
to order ${\cal O}\left (n^{-1}\right )$, by using the expressions obtained through Cox and Snell's (1968)
formulae. We also considered two schemes of bias correction based
on bootstrap. 

Finally, we considered a simulation study in a nonlinear reciprocal gamma model with nonlinear dispersion covariates. The simulation suggested, among other things, that bias-corrected up to the second-order estimators should be used instead of the usual MLEs. Furthermore, we were able to notice that the analytical bias-corrected estimators had the smallest biases, whereas the bias-corrected estimators using parmetric bootstrap scheme had the smallest mean square error. Note that, even though the parametric bootstrap had the least mean square error, this fact yielded that the confidence intervals induced by the bootstrap estimator had the poorest coverage, mainly because its small variance produced confidence intervals with small length. Nevertheless, the confidence intervals obtained by the corrective method were the best in terms of coverage closer to the nominal level.

\section*{Appendix}
We give explicit expressions for the cumulants and their derivatives, both defined in Section 3. Further, we give the expressions for each
quantity contained in equations (\ref{biasbeta}) and \eqref{biastheta}, some of them are also deduced to help the reader who might be interested in checking the results.

Consider initially the following notation for the derivatives, and product of the derivatives, of the predictor with respect to the regression parameters:
$$(rs)_i = \frac{\partial^2\eta_{1i}}{\partial\beta_r\partial\beta_s},\quad (RS)_i = \frac{\partial^2\eta_{2i}}{\partial\theta_R\partial\theta_S},\quad (rs,T)_i = \frac{\partial^2\eta_{1i}}{\partial\beta_r\partial\beta_s}\frac{\partial\eta_{2i}}{\partial\theta_T},$$
and so on. Recall that
$$d_{ri} = E\left[\frac{\partial^r}{\partial\mu_i^r} t(Y_i,\mu_i)\right],\quad\hbox{and}\quad \alpha_{ri} =E\left[\frac{\partial^r}{\partial\phi_i^r}a(\phi_i,Y_i)\right].$$

By using these quantities, the cumulants can be written as
\begin{eqnarray*}
\kappa_{rs} &=& \sum_{i=1}^n \phi_i d_{2i} \left(\frac{d\mu_i}{d\eta_{1i}}\right)^2 (r,s)_i,\\
\kappa_{rS} &=& 0,\\
\kappa_{RS} &=& \sum_{i=1}^n \alpha_{2i} \left(\frac{d\phi_i}{d\eta_{2i}}\right)^2 (R,S)_i,
\end{eqnarray*}
\begin{eqnarray*}
\kappa_{rsu} &=& \sum_{i=1}^n\phi_i\left\{ d_{3i} \left(\frac{d\mu_i}{d\eta_{1i}}\right)^3 + 3 d_{2i} \frac{d\mu_i}{d\eta_{1i}}\frac{d^2\mu_i}{d\eta_{1i}^2}\right\} (r,s,u)_i\\
&&+ \sum_{i=1}^n \phi_i d_{2i} \left( \frac{d\mu_i}{d\eta_{1i}}\right)^2 \{ (rs,u)_i + (ru,s)_i + (su,r)_i\},\\
\kappa_{rsU} &=& \sum_{i=1}^n d_{2i}\left(\frac{d\mu_i}{d\eta_{1i}}\right)^2\frac{d\phi_i}{d\eta_{2i}}(r,s,U)_i,\\
\end{eqnarray*}
\begin{eqnarray*}
\kappa_{rSU} &=&0,\\
\kappa_{RSU} &=& \sum_{i=1}^n \left\{ \alpha_{3i} \left(\frac{d\phi_i}{d\eta_{2i}}\right)^3 + 3 \alpha_{2i} \frac{d^2\phi_i}{d\eta_{2i}^2}\frac{d\phi_i}{d\eta_{2i}}\right\}(R,S,U)_i\\
&&+\sum_{i=1}^n \alpha_{2i} \left(\frac{d\phi_i}{d\eta_{2i}}\right)^2 \{ (RS,U)_i + (RU,S)_i + (SU,R)_i\}.
\end{eqnarray*}

Differentiating the second order cumulants with respect to the parameters, we have
\begin{eqnarray*}
\kappa_{rs}^{(u)} &=& \sum_{i=1}^n \phi_i \left\{d_{2i}'\left(\frac{d\mu_i}{d\eta_{1i}}\right)^3 + 2d_{2i} \frac{d\mu_i}{d\eta_{1i}} \frac{d^2\mu_i}{d\eta_{1i}^2}\right\} (r,s,u)_i\\
&&+ \sum_{i=1}^n \phi_i d_{2i} \left(\frac{d\mu_i}{d\eta_{1i}}\right)^2 \{(ru,s)_i + (su,r)_i\},\\
\kappa_{rs}^{(U)} &=&  \sum_{i=1}^n d_{2i} \left(\frac{d\mu_i}{d\eta_{1i}}\right)^2 \frac{d\phi_i}{d\eta_{2i}} (r,s,U)_i,\\
\kappa_{RS}^{(u)} &=& 0,\\
\kappa_{RS}^{(U)} &=& \sum_{i=1}^n \left\{ \alpha_{2i}' \left(\frac{d\phi_i}{d\eta_{2i}} \right)^3 + 2 \alpha_{2i} \frac{d\phi_i}{d\eta_{2i}} \frac{d^2\phi_i}{d\eta_{2i}^2}\right\}(R,S,U)_i\\
&&+ \sum_{i=1}^n \alpha_{2i} \left(\frac{d\phi_i}{\eta_{2i}} \right)^2 \{ (RU,S)_i + (SU,R)_i \},
\end{eqnarray*}
\begin{eqnarray*}
\kappa_{rS}^{(u)} &=& 0,
\end{eqnarray*}
\begin{eqnarray*}
\kappa_{rS}^{(U)} &=& 0.
\end{eqnarray*}

We now recall let $M_1$, $M_2$ and $M_3$ be the diagonal matrices given in equations \eqref{m1} and \eqref{m2m3}. Let $m_{ji}$ be the $i$th diagonal element of the matrix $M_j$.  Also, let $W_\beta = {\rm diag} \left(-d_{2i} (d\mu_i/d\eta_{1i})^2\right)$ and $W_\theta = {\rm diag} \left(-\alpha_{2i} (d\phi_i/d\eta_{2i})^2\right)$, and $w_{bi}$, and $w_{ti}$ be the diagonal elements of $W_\beta$ and $W_\theta$, respectively. We then, have that the ${\cal O}(n^{-1})$ bias of $\hat\beta$, $B(\hat\beta)$ is
\begin{eqnarray*}
B(\hat\beta_a) = \sum_{r,s,u} \kappa^{ar} \kappa^{su}\left\{ \kappa_{rs}^{(u)} - \frac{1}{2}\kappa_{rsu}\right\} &=& \sum_{i=1}^n \phi_im_{1i} \sum_{r} \kappa^{ar} (r)_i\sum_{s,u} \kappa^{su} (s,u)_i \\
&&- \sum_{i=1}^n \frac{1}{2}\phi_iw_{bi} \sum_{r,s,u} \kappa^{ar} \kappa^{su} \{(rs,u)_i - (ru,s)_i-(su,r)_i\}\\
&=&\sum_{i=1}^n \phi_i m_{1i} \sum_r \kappa^{ar} (r)_i \delta_i^T(\tilde{X} K^{\beta} \tilde{X}^T)\delta_i  \\
&&- \sum_{i=1}^n \frac{1}{2}\phi_i w_{bi} \sum_r \kappa^{ar} (r)_i \sum_{s,u} \kappa^{su} (su)_i\\
&=&\delta_a^T\sum_{i=1} K^{\beta}\tilde{X}^T\delta_i \phi_i m_{1i} \delta_i^T (\tilde{X} K^{\beta} \tilde{X}^T)\delta_i\\
&&- \delta_a^T\sum_{i=1}^n \frac{1}{2}\phi_i w_{bi} K^{\beta}\tilde{X}_i \delta_i E_i\\
&=& \delta_a^T K^{\beta} \tilde{X}^T\Phi M_1 Z_{\beta} - \frac{1}{2}\delta_a^T K^{\beta} \tilde{X}^T\Phi W_\beta E\mathbf{1},
\end{eqnarray*}
where $\delta_a$ is a $p\times 1$ vector with a one in the $a$th position and zeros elsewhere, and the matrices $E$, $Z_\beta$, and $K^\beta$ were defined in Section 3. 

Analogously, one uses the expression
$$B(\hat\theta_a) = \sum_{R,s,u} \kappa^{aR}\kappa^{su} \left\{\kappa_{Rs}^{(u)} - \frac{1}{2}\kappa_{Rsu}\right\} + \sum_{R,S,U} \kappa^{aR}\kappa^{SU}\left\{\kappa_{RS}^{(U)}-\frac{1}{2}\kappa_{RSU}\right\},$$
to obtain that
$$B(\hat\theta_a) = \delta_a^T K^\theta \tilde{Z}^T \{M_2 Z_\theta  - M_3 Z_\beta \} - \frac{1}{2} \delta_a^T K^\theta \tilde{Z}^T W_\theta F\mathbf{1},$$
where, in this case, $a=1,\ldots,q$, and the matrices $Z_\beta$, $K^\theta$, and $F$ were also defined in Section 3.

\end{document}